\documentstyle[11pt,fleqn]{article}
\def\titolo{\par\bigskip\begin{center}\bf\LARGE}
\def\endtitolo{\end{center}\par\bigskip\par\rm\normalsize}
\def\instit{\begin{center}\large}
\def\endinstit{\end{center}\rm\normalsize}
\def\references{\begin{thebibliography}{10}}
\def\endreferences{\end{thebibliography}}

\newcommand{\btit}{\begin{titolo}}
\newcommand{\etit}{\end{titolo}}

\renewcommand{\author}[1]{\begin{center}\Large #1\end{center}}
\renewcommand{\date}[1]{\par\bigskip\par\sl\hfill #1\par\medskip\par\rm}

\newcommand{\pacs}[1]{\smallskip\noindent{\sl PACS number(s):
                       \hspace{0.3cm}#1}\par\bigskip\rm}
\newcommand{\babs}{\hrule\par\begin{description}\item{Abstract: }\it}
\newcommand{\eabs}{\par\end{description}\hrule\par\medskip\rm}

\newcommand{\hs}{\qquad\qquad}         
\newcommand{\nn}{\nonumber}            
\newcommand{\beq}{\begin{eqnarray}}    
\newcommand{\eeq}{\end{eqnarray}}      
\newcommand{\beqn}{\begin{eqnarray}}   
\newcommand{\eeqn}{\end{eqnarray}}     
\newcommand{\fr}[2]{\mbox{$\frac{#1}{#2}$}}      
\newcommand{\Tr}{\,\mbox{Tr}\,}                  
\newcommand{\al}{\alpha}

\newcommand{\si}{\sigma}

\begin{document}

\thispagestyle{empty}
\renewcommand{\thefootnote}{\dagger}

KEK-TH-397 - KEK Preprint 94-33

\date{May, 1994}
\vskip 2truecm

\begin{center}
{\bf Interaction of Low - Energy Induced Gravity with Quantized Matter \\
and \\
Phase Transition Induced by Curvature}
\vskip 2truecm

\bigskip
\renewcommand{\thefootnote}{\ddagger}

{\bf I. L. Shapiro\footnote { e-mail address:
shapiro@fusion.sci.hiroshima-u.ac.jp}}

\bigskip

{\sl Tomsk State Pedagogical Institute, Tomsk, 634041, Russia \\
and \\
Department of Physics, Hiroshima University \\
Higashi - Hiroshima, Hiroshima, 724, Japan}

\bigskip

{\bf G. Cognola\footnote {e-mail address:
cognola@science.unitn.it}}

\bigskip

{\sl Dipartimento di Fisica, Universit\`a di Trento\\
and\\ Istituto Nazionale di Fisica Nucleare, Gruppo Collegato di
Trento, Italia}

\end{center}
\vskip 2.5truecm

\noindent
{\bf Abstract.} At high energy scale the only quantum effect of any asymptotic
free and asymptotically conformal invariant GUT is the trace anomaly of the
energy - momentum tensor. Anomaly generates the new degree of freedom,
that is propagating conformal factor.
At lower energies conformal factor starts to interact with scalar field
because of the violation of conformal invariance.
We estimate the effect of such an interaction and find the running of the
nonminimal coupling from conformal value $\frac{1}{6}$ to $0$.
Then we discuss the possibility of the first order phase transition
induced by curvature in a region close to the stable
fixed point and calculate the induced values of Newtonian and cosmological
constants.


\setcounter{page}1
\renewcommand{\thefootnote}{\arabic{footnote}}
\setcounter{footnote}0
\newpage

\section{Introduction}

The cosmological constant problem remains the more misterious one in a modern
high energy physics. There was proposed a number of different
approaches to solve this problem (see, for example, [1 - 11] and
references therein), but no any approach is able to give the
completely consistent scheme of vanishing this constant.

Here we shall consider the influence of vacuum quantum effects of matter
fields in curved space-time to the value of induced cosmological constant.
In a region of asymptotic freedom, that is,
in an accord with the modern point of view, at energies beyond
the Grand Unification scale, all interactions between
matter fields are weakened and the only quantum effect is the appearance of
the trace anomaly of the energy-momentum tensor.
The purpose of the present paper is to explore the back reaction
of this vacuum effect to the matter fields
with respect to induced value of the cosmological
constant. Trace anomaly generates a new dynamical degree of freedom, which is
usually named as conformal factor or dilaton. Not so far
Antoniadis and Mottola have considered the theory of conformal factor as some
infrared version of quantum gravity [12], and found this theory as a useful
tool for the exploration of the cosmological constant problem.
Then this approach (with various modifications) has been developed in a
few papers (see, for example [13 - 20]).
In particular, in Ref.~[20] there was calculated the contributions
of quantized conformal factor to the effective potential of scalar field.
At the same time in Ref.~[20] the semiclassical approximation
has been used and quantum effects of the matter fields was not taken into
account. The complete investigation of the system of interacting matter and
dilaton field meets some difficulties, because in the corresponding
theory there are fourth as well as second derivative terms. Here we consider
the complete case and derive one-loop divergences with the use of method
proposed in [21] (see also [22]). Then we point out the one-loop
renormalizability of the theory and use renormalization group methods to
analyze the running of the coupling and to derive the effective potential of
the scalar field. After this we consider the possibility of
the first order phase transition induced by curvature, and discuss a possible
way to fine tune the scale parameter in order to provide a
small value of the induced cosmological constant.

The paper is organized as follows. In section 2 we give the overview of
the basical concepts of [20] and formulate the structure of interaction between
the conformal factor and matter fields.
Section 3 is devoted to the calculation of the one-loop counterterms and to the
renormalization structure. In section 4 the renormalization group method is
applied for the analysis of the running of coupling constants. Here we also
obtain the expression for the effective potential.
In section 5 the possibility of the first
order phase transition is stated and the induced values of Newtonian and
cosmological constants are calculated.

\section{The action of conformal factor and coupling structure}

The starting point of our investigation is the theory of
asymptotically free  massless fields
of spin $0$, ${1\over2}$ and $1$ in an external
gravitational field.
One can find the review of quantum field theory in an extermal gravitational
field for example in Refs.~[23,24,22].
In particular, in [22] it is also presented the theory
of interacting fields in curved space-time. The multiplicative
renormalizability require the nonminimal terms to be included into the action
as well as the vacuum ones. When the radiational
corrections are taken into account, the parameter
of nonminimal coupling obeys the corresponding
renormalization - group equations.
As it was pointed out in [25,26] (see also [22]),
in some asymptotically free models it takes place the asymptotic
conformal invariance.
This means that the nonminimal coupling $\xi$
(we suppose the nonminimal term to have the form $\xi R\phi^2$)
is arbitrary at low energies while at high energies it has the
conformal value $\frac{1}{6}$.

The next important quantum effect, in an external gravitational field,
is the appearance of the anomaly trace
of the energy-momentum tensor which allows to calculate, with accuracy to
some conformal invariant functional, the effective action of vacuum [27,28]
(see also [22]).
This effective action originally arises as a nonlocal functional,
but it can be written in a local form with the help
of an extra dimensionless field, which is named as dilaton,
in analogy with the string theory, or as conformal
factor.

The anomaly trace of the energy-momentum tensor has the form [29,23,24]
\beq
T=<T_{\mu}^{\;\;\mu}>=k_1C^2+k_2E+k_3\Box R
\label{1}\:,\eeq
where the values of $k_1,_2,_3$ are determined by the
number of fields of different spin in a starting GUT model. Trace anomaly
(1) leads to
the equation for the effective action
\beq
-{2\over\sqrt{-g}}
g_{\mu\nu}{\delta W \over{\delta g_{\mu\nu}}}=T
\:,\nn\eeq
which has the nonlocal solution [27,28]
\beq
W[g_{\mu\nu}]&=&
S_c[g_{\mu\nu}]+\int d^4x\sqrt{-g}
\left(k_3+\fr23k_2\right)R^2
+\int d^4x\sqrt{-g_x}\int d^4y\sqrt{-g_y}
\nn\\&&\hs
\times\;\left\{k_1C^2+\fr12k_2
\left(E-\fr23\Box R\right)\right\}_x
G(x,y)\left\{k_2\left(E-\fr23\Box R\right)
\right\}_y
\label{2}\:,\eeq
where $G(x,y)$ is the Green function for
the Hermitian conformal covariant fourth-order operator (4),
$C^2$ is the square of the Weyl tensor and $E$ is Gauss-Bonnet invariant.
This effective action can be written in a local form with the help of
an auxiliary dimensionless field $\sigma$ [27]. It reads
\beq
W[g_{\mu\nu},\sigma] = S_{c}[g_{\mu\nu}]+
\int d^4x\sqrt{-g}\left\{\fr12\sigma\Delta\sigma
+\sigma\left[k'_1C^2+k'_2\left(E-\fr23\Box R\right)\right]
+k'_3R^2\right\}
\label{3}\:.\eeq
The Conformally covariant self-adjoint operator $\Delta$
is defined by
\beq
\Delta=\Box^{2}+
2R^{\mu\nu}\nabla_{\mu}\nabla_{\nu}-\fr23R\Box
+\fr13(\nabla^{\mu}R)\nabla_{\mu}
\label{4}\:.\eeq
The values of $k'_{1,2,3}$ differ from $k_{1,2,3}$ because of contribution
of $\Delta$ to conformal trace (1) [27].
The solution (3) contains an arbitrary conformal invariant
funtional $S_{c}$ , which is the integration constant for the
equation (2). This functional is not essential for our purposes and we
shall not take it into account.

Our main supposition is that the quantum effects of induced gravity,
that is of the field $\sigma$,
are relevant below the scale of asymptotic freedom
and asymptotic conformal invariance, where coupling constants in the matter
fields sector are not equal to zero. We are interested in the cosmological
applications, and therefore it is natural to suppose that the transition to
low energies (or long distances) corresponds to some conformal transformation
in the induced gravity action (3) and hence classical fields and
induced gravity appear in different conformal points [32].
To realize this it is necessary
to make a conformal transformation of the metric in (3) and then to consider
the unified theory. At the same time it is more convenient to make the
conformal transformation of metric and matter fields in the action of the last.
Such a transformation corresponds to some change of variables in the path
integral for the unified theory.

The only source of conformal noninvariance in the action of the fields of
spin $0,\frac{1}{2},1$ is the nonminimal term in the scalar sector.
In the framework of asymptotically conformal invariant models the value of
$\xi$ is not equal to $\frac{1}{6}$ at low energies
and hence the interaction of conformal factor with scalar field arises.
Introducing the scale parameter $\alpha$ we obtain the following action
for the conformal factor coupled with the scalar field:
\beq
S&=&W[g_{\mu\nu},\sigma]+\int d^{4}x \sqrt{-g}
\left\{
\fr12(1-6\xi)\phi^2\left(\alpha^2
g^{\mu\nu}\partial_{\mu}\sigma\partial_{\nu}\sigma
+\alpha\Box\sigma\right)
\right.\nn\\&&\hs\hs\hs\left.
+\fr12g^{\mu\nu}\partial_{\mu}\phi\partial_{\nu}\phi
+\fr12\xi R\phi^2-\fr1{24}f\phi^4\right\}
\label{5}\:,\eeq
where $W[g_{\mu\nu},\sigma]$ is defined in [3].
Thus the interaction between scalar field and conformal factor
arises as a result of the conformal transformation of the metric
$g_{\mu\nu}\to g'_{\mu\nu}=g_{\mu\nu}\exp(2\alpha\sigma)$
and the matter field
$\Phi\rightarrow \Phi'=\Phi\exp(d_{\Phi}\alpha\sigma)$,
where $d_{\Phi}$ is the conformal weight of the field
$\Phi$. The only kind of fields which takes part in such an interaction is
the scalar one,
where the interaction with conformal factor appears as a result of
nonconformal coupling at low energies.
Hence the contributions of other matter fields to the
effective potential of the scalar field do not depend on conformal
factor and can be calculated separately. In the next sections we shall
concentrate ourselves on the theory (5) and estimate the dilaton contributions
to the effective potential.

\section{Calculation of one-loop divergences}

The calculation of the one-loop divergences of the theory (5) is not a
trivial problem because the classical action
contains second derivative terms as well
as fourth derivative ones. Here we shall use the method of Ref. [21], where
the one-loop divergences have been calculated for higher derivative
quantum gravity coupled with matter fields.
According to [21], we shall use the background field method and hence
we start with the separation of fields into background
$\sigma,\phi$ and quantum $\tau,\eta$ ones, by means
\beq
\sigma\rightarrow \sigma' = \sigma + \tau \:,\hs
\phi \rightarrow \phi' = \phi + \eta
\label{6}\:.\eeq

The one-loop effective action is defined as
\beq
\Gamma = \frac{i}{2} \Tr\ln {\hat{H}}
\label{7}\:,\eeq
where ${\hat{H}}$ is the bilinear
(with respect to quantum fields $\tau,\eta$) form of the classical action (5).
After some algebra we get the following self-adjoint bilinear form
\beq
\hat{H}=\left(\matrix{
H_{\tau\tau} &H_{\tau\eta}\cr
H_{\eta\tau} &H_{\eta\eta}\cr} \right)
\label{8}\:,\eeq
where
\beq
H_{\tau\tau} = \Box^{2}+
2R^{\mu\nu}\nabla_{\mu}\nabla_{\nu}-\fr23R\Box
+\fr13(\nabla^{\mu}R)\nabla_{\mu}
+(6\xi-1)\left[\alpha^2\phi^2\Box
+\alpha^2(\nabla^{\mu}\phi^2)\nabla_{\mu}\right]
\:,\nn\eeq
\beq
H_{\tau\eta} =
-(6\xi-1)\left[\al\phi\Box+2\al(\nabla^\mu\phi)\nabla_\mu
+\al(\Box\phi)
-2\al^2\phi(\nabla^\mu\si)\nabla_\mu
-2\al^2(\nabla_\mu(\phi\nabla^\mu\si))\right]
\:,\nn\eeq
\beq
H_{\eta\tau}=-(6\xi-1)\left[\al\phi\Box
+2\al^2\phi(\nabla^\mu\si)\nabla_\mu\right]
\:,\nn\eeq
\beq
H_{\eta\eta} =-\Box+\xi R-\fr12f\phi^2
-(6\xi-1)\left[\alpha(\Box\sigma)
+\alpha^2(\nabla^\mu\sigma)
(\nabla_\mu\sigma)\right]\:.
\label{9}\eeq
After the change of variables $\eta\rightarrow i\eta$ we arise to the
following structure of $\hat{H}$
\beq
\hat{H} =\left(\matrix{
\Box^{2}+2V^{\mu\nu}\nabla_{\mu}\nabla_{\nu}+N^{\mu}\nabla_{\mu}+U
&\qquad Q_1\Box + Q_2^{\mu}\nabla_{\mu} + Q_3\cr
P_1\Box + P_2^{\mu}\nabla_{\mu} + P_3 &\qquad\Box+E^\mu\nabla_\mu+D
}\right)
\label{10}\:,\eeq
where the values of $V,N,U,Q_i,P_i,E,D$ follows from (9). They read
\beq
V^{\mu\nu} =
2R^{\mu\nu}-\fr23Rg^{\mu\nu}+
(6\xi - 1)\phi^2\alpha^2g^{\mu\nu}
\:,\nn\eeq
\beq
N^{\mu}=\fr13(\nabla^{\mu}R)
+(6\xi-1)\alpha^2(\nabla^{\mu}\phi^2)
\:,\hs U=0
\:,\nn\eeq
\beq
Q_1=-i(6\xi - 1)\phi \alpha
\:,\hs
Q_2^\mu=- 2i(6\xi - 1)\left[(\nabla^\mu\phi)\alpha
-\phi(\nabla^\mu\sigma)\alpha^2\right]
\:,\nn\eeq
\beq
Q_3=-i(6\xi-1)\left[\alpha(\Box\phi)
-2\al^2(\nabla_\mu(\phi\nabla^\mu\si))\right]
\:,\hs
P_1=-i(6\xi - 1)\phi \alpha
\:,\nn\eeq
\beq
P_2^\mu=-2i(6\xi - 1)\phi (\nabla^\mu\sigma)\alpha^2
\:,\hs  P_3=0
\:,\hs E^\mu=0\:,\nn\eeq
\beq
D=-\xi R+\fr{1}{2}f\phi^2
+(6\xi-1)\left[\alpha(\Box\sigma)+
\alpha^2(\nabla^\mu\sigma)(\nabla_\mu\sigma)\right]\:.
\label{11}\eeq

Expressions (10) enables us to use the method and partially also the
results of [21] in a direct way.
After some algebra we obtain the following
general form for the one-loop divergences of the effective action (7):
\beq
\Gamma_{div}&=&\frac2\varepsilon
\int d^4x\sqrt{-g}\left\{
\fr14P_2^\mu
Q_{2\mu}+\fr14P_1 Q_3-\fr14V^{\mu}_{\mu}P_1 Q_1
\right.\nn\\&&\hs
-DP_1 Q_1+\fr12{(P_1 Q_1)}^2
+\fr12Q_2^\mu\nabla_\mu P_1-\fr16RP_1 Q_1
\nn\\&&\hs
+\fr1{24}V^{\mu\nu}V_{\mu\nu}+\fr1{48}(V^{\mu}_{\mu})^2
-\fr16V^{\mu\nu}R_{\mu\nu}+\fr1{12}V^\mu_\mu R-U
\nn\\&&\hs\left.
+\fr12{\left(D+\fr16R\right)}^2
+\fr1{20}\left(R^{\mu\nu}R_{\mu\nu}-\fr13R^2\right)
-\fr1{36}R^2\right\}
\nn\\&&\hs\hs
+\mbox{ (surface terms)}
\label{12}\:.\eeq
Substituting (11) into (12) we arise at the explicit expression for
$\Gamma_{div}$
\beq
\Gamma_{div}&=&\frac{2}{\varepsilon}\int d^{4}x\sqrt{-g}
\left\{
\fr1{24}\phi^4\left[3f^2+12(6\xi-1)^2f\alpha^2
-432(6\xi-1)^2\alpha^4\xi^2\right]
\right.\nn\\&&\hs
-\fr12\phi^2\left(R-6\alpha\Box\sigma
-6\alpha^2\nabla^\mu\sigma\nabla_\mu\sigma\right)
\left[\fr16(6\xi-1)f+2\alpha^2\xi(6\xi-1)^2\right]
\nn\\&&\hs\left.
-\fr7{60}\left(R^{\mu\nu}R_{\mu\nu}-\fr13R^2\right)
+\fr1{72}\left(R-6\alpha\Box\sigma
-6\alpha^2\nabla^\mu\sigma\nabla_\mu\sigma\right)^2
\right\}
\nn\\&&\hs\hs\hs
+\mbox{ (surface terms)}
\label{13}\:.\eeq
where $\varepsilon = (4\pi)^2(n - 4)$ is parameter of dimensional
regularization.

Let us now make some notes concerning the renormalization structure in
the theory under consideration.
One can easily see that in this theory the wide
cancellation of divergences takes place. This happens because such a
theory is renormalizable, at least at one-loop level,
if the necessary vacuum terms are introduced.
The needed form of the vacuum action is clear from (13) and is
\beq
S_{vac}=\int d^4x \sqrt{-g}\left\{
a_1\left(R-6\alpha\Box\sigma
-6\alpha^2\nabla^\mu\sigma\nabla_\mu\sigma\right)^2
+a_2\left(R^{\mu\nu}R_{\mu\nu}-\fr13R^2\right)
\right\}
\label{14}\:.\eeq

Note that the last action contains the dynamical field $\sigma$ and therefore
can not formally be viewed as vacuum one. At the same time
the dependence of $\sigma$ can be removed by conformal transformation
which have to be included into the renormalization.
Another way need the terms of the action (14) to be included from the very
beginning. In this case the consideration presented below corresponds
to the more complete theory taken in the stable fixed point for the
corresponding effective couplings [19, 41].

Next point is related to the renormalization in the sector of scalar
field $\phi$. One can see that there is no any divergences
which lead to renormalization of $\phi$. All the divergences
can be removed by the renormalization of couplings, including $a_1,a_2$
and the renormalization group equations include only $\beta$, but not $\gamma$
functions. Then it follows that the
effective potential of the theory depends on $\beta$-functions only, and
therefore it is free of ambiguities which arise, for instance,
in gauge theories (see, for example, discussion in [30]
for the case of higher derivative
gravity corrections to effective potential).

\section{Renormalization group equations}

Since the theory defined by Eq.~(5) is multiplicatively renormalizable,
one can use the renormalization group method for its study.
For our purposes it is more convenient to deal with
an arbitrary background metric $g_{\mu\nu}$ and therefore
we must use the approach described in [22]
(see also original papers [25,26,31]).
The general solution of the renormalization group
equations for effective action
\beq
\left\{ \mu\frac{d}{d\mu}+\beta_f \frac{d}{d f}
+\beta_\xi \frac{d}{d \xi}
-\gamma_{\phi}\frac{\delta}{\delta \phi}
- \gamma_{\sigma}\frac{\delta}{\delta \sigma}
\right\}\Gamma[\phi,\sigma,f,\xi,g_{\mu\nu},\mu]=0
\label{15}\eeq
has the form
\beq
\Gamma[\phi, \sigma,f,\xi,g_{\mu\nu}e^{2t}, \mu] =
\Gamma [\phi(t),\sigma(t),f(t),\xi(t),g_{\mu\nu},\mu]
\label{16}\:,\eeq
where $\mu$ is the dimensional parameter of renormalization.
Effective fields and coupling constants obey the equations
\beq
\frac{d\phi(t)}{dt} = (\gamma_{\phi} + 1)\phi\:,\hs\phi(0)=\phi\:,
\nn\eeq
\beq
\frac{d\sigma(t)}{dt}=\gamma_{\sigma}\sigma,
\hs \sigma(0) = \sigma\:,
\nn\eeq
\beq
\frac{df(t)}{dt'} = \beta_f,
\hs f(0) = f\:,
\nn\eeq
\beq
\frac{d\xi(t)}{dt'} = \beta_\xi,
\hs \xi(0) = \xi \:.
\label{17}\eeq
Here $t' = (4\pi)^{- 2}t$, while $\gamma$ and $\beta$
functions are defined as usual. Note that in our case
there is no need to renormalize fields $\phi, \sigma$ and therefore all
$\gamma$ functions are equal to zero. The $\beta$ - functions for
effective couplings $f(t), \zeta(t) = 1-\xi(t)$
can be easily obtained from (13). One has
\beq
(4\pi)^2\beta_f = 3f^2 + 12f\alpha^2 \zeta^2 + 12\alpha^4 \zeta^2
{(\zeta - 1)}^2\:,\hs f(0) = f\:,
\nn\eeq
\beq
(4\pi)^2\beta_\zeta = \zeta [f + 2\alpha^2 \zeta (\zeta - 1)]\:,
\hs \zeta(0) = \zeta\:.
\label{18}\eeq
Here we introduce $\zeta$ instead of $\xi$ for compactness.
Let us now consider the asymptotics of the effective couplings
$f(t), \zeta(t)$. It is easy to see that both $\beta$ functions (18) vanish
in the physically relevant points $f = 0, \zeta = 0$ and $f = 0, \zeta = 1$.
There are also three more solutions with negative $f$, but they do not
look so interesting, because classical potential for $\phi$
in these cases is not bounded from below.
The first solution corresponds to the conformal fixed point while
the second one corresponds to the minimal fixed point.
The last means that within this solution $\xi$ is
equal to zero. One can easily make the analysis of stability of
the minimal fixed point in the framework of standard Lyapunov's method and
find it stable in the IR limit $t \rightarrow - \infty$.
Moreover one can easily obtain, that in this limit
$f\sim\xi^6\rightarrow0$ and therefore $f\ll\xi$.

The problem of stability of the first solution can not be solved in such a
way, because if we make the infinitesimal variations of couplings
$\zeta,f$, the linear corrections to this fixed point in $\beta$ functions
are equal to zero.
Rejecting the infinitesimal terms of higher order we find
\beq
\frac{d\delta f}{dt'}=3(\delta f)^2
+12\alpha^4(\delta\zeta)^2\:,
\nn\eeq
\beq
\frac{d\delta\zeta}{dt'}
=\delta\zeta(\delta f+2\alpha^2\delta\zeta)\:,
\label{19}\eeq
where $\delta\zeta,\delta f$ are infinitesimal variations of couplings.
Since we are only interested in the infinitesimal variations, there are three
possible cases: i)  $\delta\zeta$ and $\delta f$ are of the same order,
ii) $\delta\zeta$ is smaller than $\delta f$, iii)$\delta f$
is smaller than $\delta\zeta$.
A detailed analysis shows that all these three
situations are not possible, and therefore conformal fixed point is not stable
in IR limit. Moreover one can easily see that both fixed points are not stable
in UV limit.

Some notes are in order. We start our consideration with some
asymptotically free and asymptotically conformal invariant GUT, and consider
the scalar field $\phi$ as part of this GUT. Formally the contribution
of quantum field $\sigma$ to $\beta_f$ contradicts asymptotic freedom and
therefore the whole approach looks quite ambiguous,
but it is not the case.
We suppose that the value of $\alpha$ to be close to zero
and hence the contributions
of quantum conformal factor are small at short distances.
On the other hand, at long distances, the scaling parameter
$\alpha$ has some finite value and this
leads to some nontrivial dynamics in far IR. Note that the IR dynamics of
$\alpha$ has been recently studied in [33].

\section{Effective potential for scalar field and phase transition induced
by curvature}

Now we follow [31,22] and use the renormalization group method for
the derivation of the effective potential for scalar field $\phi$.
According to [31,22] the solution of Eq.~(15) for the potential part of
the effective action has the form
\beq
V_{eff}=-\frac12\xi R\phi^2+\frac{f}{24}\phi^4
+\frac1{48}\beta_f\phi^4\left[\ln\frac{\phi^2}{\mu^2}
-\frac{25}6\right]
-\frac14\beta_\xi\phi^2 R\left[\ln\frac{\phi^2}{\mu^2}-3\right]
\label{20}\:,\eeq
where $\mu$ is a dimensional parameter of renormalization and both $\beta$
functions have been defined in Eq.~(18).

Here we shall restrict ourselves to only consider
the first-order phase transition.
Then the equations for the critical values of curvature
$R_c$ and order parameter $\phi_c$ have the form
\beq
V_{eff}(R_c,\phi_c) = 0, \hs V'_{eff}(R_c,\phi_c) = 0,
\hs V''_{eff}(R_c,\phi_c) > 0.
\label{21}\eeq
Here primes stand for derivatives of $V$ with respect to $\phi$.
The effective potential in Eq.~(20) has rather complicated form and therefore
any relevant analysis of Eqs.(21) needs some restrictions on the
values of $f,\zeta$ to be imposed.
Since we are interested in the phase transition,
which has to take place in far infrared, it is natural to suppose
that these couplings obey the renormalization group equations and have values
which are close to the IR stable fixed point $f=0,\xi=0$.

Then the first two equations in (21) have two non trivial
solutions for critical values of curvature and $\phi$. They read
\beq
\phi_c^2 = q_{1,2}|R_c|, \hs
q_{1} = - \frac{\varepsilon}{18 \alpha^4 \xi^4}, \hs
q_{2} = - \frac{\varepsilon}{2}\:,
\nn\eeq
\beq
(\phi_c^{(1)})^2=\mu^2\exp\left(\frac{18\alpha^2+1}{6\alpha^2}\right)
\:,\hs
(\phi_c^{(2)})^2=\mu^2\exp\left(\frac{25}{6}\right)\:,
\label{22}\eeq
where $\phi_c^{(1,2)}$ correspond to $q_{1,2}$ and
$\varepsilon=R/|R|$ is the sign of the scalar curvature.
For $q_{2}$ the third condition in Eq.~(21) reads as $\xi+o(\xi^2)<0$,
which contradicts the initial condition $\xi\approx\frac16>0$,
which must hold at high energies.
For $q_{1}$ the third condition in Eq.~(21) has the form
$\frac{\beta_f}{32}+o(\xi^3)>0$ and so the initial condition is fulfilled.
Therefore in the framework of used approximation the first order phase
transition takes place at critical values of $\phi_c$ and $|R_c|$
which correspond to the choice $q_1$ in Eq.~(22).

Substituting the values of $\phi_c^{(1)}$ and $q_{1}$ from
Eq.~(22) into effective potential (20),
we obtain the estimate for $V_{eff}$ at the critical point.
The corresponding action has the form of Hilbert-Einstein action
\beq
S_{ind} = -\frac{1}{16\pi G_{ind}}\int d^{4}x\sqrt{-g}
(R - 2\Lambda_{ind}) \:,
\label{23}\eeq
where induced values of Newtonian and cosmological constants are defined
from Eqs.~(20)-(22) to be
\beq
\frac{1}{16\pi G_{ind}} =-\xi{(\phi_c^{(1)})}^2(1-2\alpha^2)^2
\:,\label{24}\eeq
\beq
\Lambda_{ind} =-\frac{1}{16\pi G_{ind}}
\frac{9 \alpha^4}{4{(1-2\alpha^2)}^2}\:.
\label{25}\eeq

If we substitute into (25) the critical value
for the order parameter $\phi^{(1)}$,
the induced Newtonian and cosmological constants are expressed via
dimensional parameter of renormalization $\mu$ and in place of
(24) we get
\beq
\frac{1}{16\pi G_{ind}} = - \xi {(1-2\alpha^2)}^2
\mu^2\exp\left( \frac{18\alpha^2+1}{6\alpha^2} \right)\:.
\label{26}\eeq
To estimate the induced values of $\Lambda_{ind}$ and
$G_{ind}$ we must remove arbitrariness related with the value of $\mu$.
There are two different ways to do this. One can follow [31,33]
and fix the induced value of Newtonian constant
to be equal to its classical value.
This means that one chooses $\mu$ of the same order of the Planck
mass $m_p =(8\pi G)^{-1}$.
At the same time we are dealing here not with
Planck energies as in Ref.~[33], but with some energy scale below the
unification point $M_x$. On the other hand we have not any ground
to suppose that the induced values of Newtonian and cosmological constants
were the same at $M_x$ scale and at the modern epoch.
On the contrary, there are some reasons in favour of the
effective running of these constants at energies above
(see, for example, [34,35]) and below [11] this scale.
Therefore it is more reasonable to choose the value of $\mu$ close to
$M_x$ or below and thus obtain the values of constants,
which are induced by the quantum effects of conformal factor
discussed above.
Note that the absence of cosmological constant in a modern observational data
needs its independent suppression in any part of the energy scale [1].
If one takes the value of $\alpha$ to be close to zero,
then the induced value of the cosmological constant will be very small.
In fact we suppose that the energy scale of strong matter
effects in external gravitational field is some
$\mu_1$ and that the quantum effects of conformal factor is relevant
at another scale $\mu_2<\mu_1$.
Since the close scales correspond to a small value of the scale
factor $\alpha$, we obtain that, if the difference
between $\mu_2$ and $\mu_1$ is small enough,
the induced value of the cosmological constant is small too.
So, if we suppose that the point of phase transition
discussed above is close to $M_x$,
then the induced value of $\Lambda$ is in a good agreement
with the observational data.

\section{Conclusion}

We have considered the interaction between conformal factor and
matter fields in a region close to the scale of asymptotic freedom.
In fact the only nontrivial contributions to $V_{eff}$ come
from the quantized scalar field.
All other fields decouple from conformal factor
and therefore give additive contributions to $V_{eff}$.
These contributions are not so essential, because in the region of
asymptotic freedom all interactions between matter fields are week.
The results of our analysis are in a good agreement with previous
semiclassical results derived in  Ref.~[20],
where the scalar field was treated as pure background.
At the same time the physical picture in the considered case is more
sophisticated. We meet here the nontrivial scale dependence of the
effective nonminimal
coupling, wich changes from conformal value $\frac{1}{6}$ to the IR stable
minimal value. The phase transition occure at the scale close to stable
fixed point, and the resulting induced constants do not depend on $\xi$,
as in the more simple case of Ref.~[20].

Now let us briefly discuss some possible ways to extend
the above considerations.
First of all it would be very interesting
to incorporate in the theory also finite temperature effects,
which have to be very important in the early Universe
\cite{kirz72-42}-\cite{albr82-48}.
Thermodynamics properties of quantum fields in static space-times
have been studied by many authors \cite{dowk78-11}.
This can be usefully done with the help of finite-temperature
generating functional in the 'imaginary time' formalism \cite{dola74-9}.
Unfortunately, in the theory we are considering one encounters
some technical difficulties due to the fact that the
small disturbance operator is of fourth order.

Next, it is possible to estimate quantum effects
of conformal factor above the unification scale.
Since this region is close to Planck energies,
it is necessary to take into account quantum effects of the metric.
To achieve renormalizability it is
necessary to consider the general dilaton gravity \cite{eliz94-5},
which is the direct generalization of higher derivative gravity
theory \cite{stel77-16}-\cite{frad82-201}.
In this case the interaction between gravity
(including conformal factor) and matter fields takes place even
without the conformal shift of the induced action.
Generally speaking, such an investigation is possible
at least at one-loop level \cite{shap94-},
but it is related with a very big volume of calculations.

\section{Acknowlegments}
 One of the authors (I.Sh.)
would like to thank Professor T.Muta for useful
discussion and also Particle Physics Group in Hiroshima University for kind
hospitality.

\end{document}